\input phyzzx    

\twelvepoint


\def\dmm{\partial_{=}}
\def\dpp{\partial_{\ne}}
\def\a{\alpha'}


\REF\Douglas{M.R. Douglas, {\it Gauge Fields and D-branes}, hep-th/9604198}
\REF\Witten{E. Witten, J. Geom. Phys. {\bf 15} (1995) 215, hep-th/9410052}
\REF\Douglastwo{M.R. Douglas, {\it Branes within Branes}, hep-th/9512077}
\REF\Vafa{C. Vafa, Nucl. Phys. {\bf B463} (1996) 435, hep-th/9512078}
\REF\Wittenthree{E. Witten, {\it On the Conformal Field Theory of the
Higgs Branch}, hep-th/9707093}
\REF\DS{D.-E. Diaconescu and N. Seiberg, {\it The Coulomb Branch of $(4,4)$
Supersymmetric Field Theories in Two Dimensions}, hep-th/9707158}
\REF\Rey{S.-J. Rey, Phys. Rev. {\bf D43} (1990) 526}
\REF\CHS{C.G. Callan, J.A. Harvey and A. Strominger, Nucl. Phys. 
{\bf B359} (1991) 611}
\REF\Lamberttwo{N.D. Lambert, Nucl. Phys. {\bf B460} (1996) 221, 
hep-th/9508039}
\REF\Wittentwo{E. Witten, Nucl. Phys. {\bf B460} (1996) 541, hep-th/9511030}
\REF\Lambert{N.D. Lambert, Nucl. Phys. {\bf B477} (1996) 141, hep-th/9605010}
\REF\CGK{E. Corrigan, P. Goddard and A. Kent, Comm. Math. Phys. 
{\bf 100} (1985) 1}
\REF\CDFN{E. Corrigan, C. Devchand, D.B. Fairlie and J. Nuyts, Nucl. Phys.
{\bf B214} (1983) 452}
\REF\HP{P.S. Howe and G. Papadopoulos, Nucl. Phys. {\bf B381} (1992) 360,
hep-th/9203070}
\REF\Reytwo{S.-J. Rey, {\it Axionic String Instantons and their Low-Energy
Implications}, UCSB-TH-89-49, in Superstrings and Particle Theory (1989);
{\it On String Theory and Axionic Strings and Instantons}, SLAC-PUB-5659, in
Particles and Fields '91 (1991)}
\REF\GPS{G.W. Gibbons, G. Papadopoulos and K.S. Stelle, 
{\it HKT and OKT Geometries on Soliton Black Hole Moduli Spaces}, 
hep-th/9706207}
\REF\VW{C. Vafa and E. Witten, Nucl. Phys. {\bf B431} (1994) 3, hep-th/9408074}
\REF\CM{C.G. Callan and J.M Maldacena, Nucl. Phys. {\bf B472} (1996) 591, 
hep-th/9602043}
\REF\SV{A. Strominger and C. Vafa, Phys. Lett. {\bf B279} (1996) 99, 
hep-th/9601029}
\REF\BvB{P.J. Braam and P. van Baal, Comm. Math. Phys. {\bf 122} (1989)
267}
\REF\HStwo{J.A. Harvey and A. Strominger, Phys. Rev. Lett. {\bf 66} (1991) 549}
\REF\Ivanova{T.A. Ivanova, Phys. Lett. {\bf B315} (1993) 277}
\REF\GN{M. G{\" u}naydin and H. Nicolai, Phys. Lett. {\bf B351} (1995) 169, 
hep-th/9502009}


\pubnum={KCL-TH-97-45\cr hep-th/9707156}
\date{July 1997}

\titlepage

\title{\bf D-brane Bound States and the Generalised ADHM Construction}

\centerline{N.D. Lambert\foot{lambert@mth.kcl.ac.uk}}

\address{Department of Mathematics\break
         King's College, London\break
         England\break
         WC2R 2LS}

\vfil

\abstract

We discuss the sigma model description of a D-string bound to $k$
D-fivebranes in type I string theory. The effective theory is an
$(0,4)$ supersymmetric  hyper-K\"ahler
with torsion sigma model on the moduli space of $Sp(k)$ instantons on 
${\bf R}^4$. 
Upon toroidal compactification to five dimensions the model is related to the 
type II picture where the target space is a symmetric 
product of $K3$'s.



\endpage


\chapter{Introduction}

In the past few years a tremendous amount of progress has been made towards
understanding the non-perturbative features of string theory. The main insights
have been obtained by studying the BPS states and in particular, BPS 
bound states.
A  central theme in the present work has been the
emergence of classical geometry as a derived concept from the underlying  
gauge theories and conformal field theories. 

D-branes are responsible for a large number of the
recent advances.  
In an elegant paper [\Douglas] Douglas showed that 
a D-string probe placed parallel to $k$ D-fivebranes in type I string theory
has Witten's ADHM massive sigma model [\Witten] 
as an effective action. The 4 massless
modes correspond to the positions of the D-string in the 
space transverse 
to the D-fivebrane and the $4k$ massive modes correspond to stretched
strings between the fivebranes and the string. To obtain the low energy 
action for the massless modes of the 
D-string (which are the worldsheet fields of a heterotic string) one must 
integrate over the 
massive modes. This leads to
an $(0,4)$ supersymmetric 
heterotic sigma model with a four dimensional target space containing  
an ADHM instanton with instanton number $k$. Thus, viewed as a probe of the 
D-fivebrane geometry, the D-string
explicitly shows the equivalence of a spacetime instanton of instanton number
$k$ with a configuration of $k$ D-fivebranes within type I string theory. 

However this picture contains another phase [\Witten,\Douglastwo]. 
When the spacetime
instanton is shrunk to zero size
another branch of the
vacuum moduli space appears with a correspondingly different low energy 
description.
If the D-string sits on the D-fivebranes then there is a branch of vacua
where the $4k$ stretched strings are massless and the 4 transverse fields 
massive.
In this phase the D-string is bound to the D-fivebrane. If we now calculate
the conformal fixed point theory describing the massless modes 
we find a heterotic sigma model with a $4k$ 
dimensional target space containing an 
``ADHM instanton of instanton number one''.\foot{Precisely what is meant by an 
ADHM instanton in 
$4k$ dimensions will be given below.}

In this paper we will focus on the bound phase of the ADHM sigma model. We will
provide an explicit description of the 
D-string/D-fivebrane
bound state of type I string theory in terms of the sigma model for 
the D-string massless modes.
Since most of the work on D-brane bound states has concentrated on
type II strings it will be helpful to see some of the details in
the type I case and compare the two. In particular for the type IIA string
on $K3$ the corresponding bound state is described by a $(4,4)$ supersymmetric 
sigma model with a symmetric product of $K3$'s as the target space. We will 
be able to relate this picture to the ADHM sigma model thus providing a 
check on type I/type II string duality
and the description of a gas of D-zerobranes [\Vafa].

In a different but not unrelated context, recent 
studies of the matrix description of M theory have involved  some  
$(4,4)$ supersymmetric analogs of the models considered here [\Wittenthree,
\DS]. In gauge theory terminology of the non-chiral $(4,4)$ models 
the probe and bound phases described here
correspond to the Coulomb and Higgs branches 
of the D-string's vacuum moduli space respectively. However there are some
crucial distinctions. In the models considered below there are no vector
multiplets and hence no Coulomb branch (the two phases are both Higgs
branches). A second distinction arises from the chiral nature of the
theories considered here. It was argued in [\DS] that there are no quantum
(i.e. $\a$) corrections on the Higgs branch of $(4,4)$ sigma models 
so that the classical solution
obtained by the hyper-K{\"a}hler quotient construction is exact. However this
argument fails for the chiral models considered here because there are 
anomalies which lead to an infinite series of $\a$ corrections to the 
metric and anti-symmetric tensor. Indeed we shall discuss the first order 
corrections in the next section.


\chapter{The Bound State Sigma Model}

We consider $k$ D-fivebranes located at $x^6=x^7=x^8=x^9=0$ with a D-string 
worldsheet placed in the $(x^0,x^1)$ plane and use the notation of 
[\Douglas]. 
D-brane quantisation rules tell
us that the massless modes (in the probe phase) of the D-string  are given 
by $(1|1)$ strings\foot{By a $(p|q)$ string we mean a string with one end on a 
D-p-brane and the other on a D-q-brane.} which 
yield 8 bosons $b^{AY},b^{AA'}$ 
and their 
superpartners
$\psi_-^{A'Y},\psi_-^{AA'}$ ($A,A',Y=1,2$). These simply 
describe the motion 
of the D-string in the space transverse and tangent to the D-fivebranes 
respectively.
Since we are in type I string theory there are also $(1|9)$ strings
which yield the fermions $\lambda_+^M$ ($M=1,2,...,32$) taking values in the
spacetime $SO(32)$ gauge bundle.
These make up the worldsheet degrees of freedom of a heterotic string.
However the presence of the D-fivebranes induces $4k$ massive modes 
$\phi^{A'm}$ (here $m=1,2,...,2k$ labels the 
fivebranes), their superpartners $\chi_-^{Am}$ and the additional fermions
$\lambda_+^{Ym}$ on the D-string.
These given by the stretched $(1|5)$ strings (with a mass proportional to the 
distance to the D-fivebranes) and introduces a potential which breaks the 
$(0,8)$ worldsheet supersymmetry to $(0,4)$.

In this paper we are only  interested in
the case of vanishing instanton size (which corresponds to setting the
$(5|9)$ strings of the D-fivebranes to zero [\Wittentwo]).
The effective action for D-string
is given by Witten's ADHM sigma model  [\Douglas,\Witten]
$$
S = S_{free} - \int\! d^2x  \left\{ 
{im\over2}\lambda_{+Ym}\phi_{B'}^{\ \ m}\psi_-^{B'Y}
+{im\over2}\lambda_{+Ym}b_{B}^{\ \ Y}\chi_-^{Bm}
+{m^2\over8}\phi^2 b^2 \right\}\ . 
\eqn\action
$$ 
Note the $b^{AA'}$ and $\psi_-^{AA'}$ fields are free because of translational
symmetry in the $x^2,x^3,x^4,x^5$ plane and the $(1|9)$ fields $\lambda_+^M$  
have decoupled in the limit of vanishing instanton size.

The probe phase corresponds to choosing the vacuum $\phi^{A'm}=0$. Here the
$b^{AY}$ fields are massless and describe the location of the D-string 
relative to the D-fivebranes. By
integrating over the massive modes one obtains an action for the massless
fields. This turns out to be a $(0,4)$ supersymmetric sigma model with a
four dimensional target space containing 
zero sized instanton
gauge field with instanton number $k$ along with gravitational corrections 
[\Witten,\Rey,\CHS,\Lamberttwo]. Furthermore the
construction of the effective action parallels the ADHM construction of 
instantons (by turning on the $(5|9)$ strings one finds the full ADHM 
construction of finite sized instantons [\Douglas,\Wittentwo]).

The bound phase corresponds to choosing the vacuum $b^{AY}=0$. In this case
the $\phi^{A'm}$ fields are massless and describe the internal state of the
D-string within the D-fivebranes. Now the $b^{AY}$ fields are massive 
indicating that the D-string has become bound to the D-fivebranes.  If we then
integrate over the massive fields to obtain the effective action for
the massless modes we find an $(0,4)$ supersymmetric sigma model with
a $4k$ dimensional target space. However, as in the probe phase a non-trivial
gauge field appears in the low energy effective action.
Furthermore the construction of this gauge field again parallels
the ADHM construction of instantons but this time 
generalised to $4k$ dimensions [\Lambert,\CGK] where we will 
find a zero-sized  instanton of instanton
number one.

Before proceeding it is necessary to clarify we mean by instantons and 
instanton
number in $4k$ dimensions. It was shown in [\CGK] that the ADHM 
construction may be generalised to $4k$ dimensions. This procedure then 
produces a gauge field whose curvature $F$ is
self-dual in the sense that 
$$
F_{A'mB'n} = \epsilon_{A'B'}F_{mn} \ ,
\eqn\Fdual
$$
with $F_{[mn]}=0$. Such a gauge field then automatically solves the Yang-Mills
field equations.
In contrast to the four dimensional case where the ADHM construction is known
to produce all self-dual gauge fields, in higher dimensions there is no
such uniqueness. This can be most easily understood by noting that above
four dimensions any notion of self-duality must break rotational 
symmetry down to some subgroup. 
There are a variety of possible subgroups [\CDFN] 
and hence a variety of notions of self-duality. However, for the purposes of 
this paper we shall only need
those configurations which are produced by this generalised ADHM construction
for which $SO(4k)$ is broken to $SU(2)\times Sp(k)$. Note also that the 
definition \Fdual\ makes no reference to the dimension and so can be 
most readily generalised.

The next point to consider
is the definition of instanton number. In four dimensions gauge fields  are
divided into topological sectors labelled by 
their first Pontryagin index. In 
general there is no such topological classification for higher dimensional 
gauge fields. However it turns out that
the instantons we shall consider may be divided into topological sectors
labelled by 
the first Pontryagin index [\CGK]
$$
p_1 = {1\over 16\pi^2}\int_{M} {\rm Tr}(F\wedge  F) \ ,
\eqn\ponty
$$
where $M$ is any of the four dimensional coordinate hyperplanes in 
${\bf R}^{4k}$. 
We therefore define the instanton number to be $p_1$ for any $k$.

To obtain the conformal fixed point action of \action\ it is helpful to
first blow up the instanton to a finite size. This is
done by adding additional interactions to \action, parameterised by a size 
$\rho$, which are 
still consistent with $(0,4)$ 
supersymmetry and  preserve the bound state (Higgs) branch of the vacuum  
(there is an essentially  unique way to
do this [\Witten]). This is analogous to the
four dimensional models found in the probe phase [\Douglas,\Lamberttwo]. One 
then looks for the 
massless fields which induce a self-dual connection (as 
constructed
by the ADHM method) to appear 
in the classical low energy effective action [\Witten,\Lambert].
Only when the size
is non-zero, and the solution is everywhere smooth can the instanton number
be easily seen to be one.
The field strength for the fixed point is then
found by taking the limit of vanishing instanton size;
$\rho\rightarrow 0$.  The generalised ADHM construction
can produce finite size instantons although it
is not clear what if anything this corresponds to physically. The only
fields we have set to zero are the $(5,9)$ strings and if  
we were to 
turn these on then the vacuum states $b^{AY}=0$ would disappear
and the bound state phase would no longer exist. Indeed we will see below 
that in
the bound phase these $(5,9)$ strings are in fact massive.

A comparison of  the field  content we are using with the definitions
in [\CGK] shows that we are producing ADHM instantons with instanton 
number one in
$4k$ dimensions with gauge group $Sp(k)$ (specifically we have 
$r=k$ and $l=1$ in the notation of [\CGK]). 
To be more explicit  we obtain 
the following gauge field strength on ${\bf R}^{4k}$ [\Lambert]
$$
F^{pq}_{mn} =
-{4\over \phi^6}
\left(
\phi^2\delta^{p}_{(m} -2\epsilon_{C'D'}\phi^{C'p}\phi^{D'}_{(m}
\right) \left( 
\phi^2\delta^{q}_{n)}-2\epsilon_{E'F'}\phi^{E'q}\phi^{F'}_{\ n)} 
\right) \ .
\eqn\Fdef
$$
The $(0,4)$ supersymmetric sigma model
conformal fixed point for the D-string in it's bound state
is  therefore (to zeroth order in $\a$)  
$$\eqalign{
S = \int\! d^2x & \left\{ 
\epsilon_{AB}\epsilon_{A'B'} \dmm b^{AA'} \dpp
b^{BB'} +i\epsilon_{AB}\epsilon_{A'B'} \psi_-^{AA'} \dpp \psi_-^{BB'}
+i\delta_{MN}\lambda_+^M \dmm \lambda^N_+
\right. \cr & \left.
+g_{A'mB'n}\dmm \phi^{A'm} \dpp \phi^{B'n}
+ig_{A'mB'n}\chi_-^{Am} \nabla^{(-)}_{\ne} \chi_-^{Bn}  
\right. \cr & \left.
+i\epsilon_{YZ}\epsilon_{mn}\lambda_+^{Ym} {\hat \nabla}_{=} \lambda^{Zn}_+
-{1\over2}\chi^{A'm}_-\chi^{B'n}_-\epsilon_{A'B'}\epsilon^{YZ}
F^{pq}_{mn}\lambda_{+Yp}\lambda_{+Zq} \right\}\ ,\cr }
\eqn\confact
$$ 
where ${\hat \nabla}$ is the covariant derivative of the gauge connection.
Due to it's chiral nature this theory is anomalous and generates a torsion
at order $\a$ which can be calculated as the Chern-Simons term for the gauge
connection. This appears in \confact\ via  
the covariant derivative with torsion $\nabla^{(-)}$. The 
model \confact\ will in addition receive higher order corrections to 
the metric and dilaton in the form of finite local counter terms need
to ensure finiteness (which is equivalent to requiring that the renormalisation
prescription preserves the extended supersymmetry [\HP]). 
To first order in $\a$ the metric is [\Lambert]\foot{The factor of $16$
corrects a factor of $12$ in [\Lamberttwo,\Lambert].}
$$
g_{A'mB'n}=
\left(1 + {16\a \over \phi^2}\right)\epsilon_{A'B'}  \epsilon_{mn}
- {16\a\over\phi^2}\epsilon_{A'B'}\epsilon_{C'D'} 
\phi^{C'}_{\ m} \phi^{D'}_{\ n} \ ,
\eqn\metricdef
$$ 
In contrast the gauge field is exact to all orders in $\a$.

Now that we have found the bound state sigma model we should  compare 
it with the description obtained from the D-fivebrane's
point of view. It's worldvolume dynamics is that of a $D=6$, $N=1$ gauge theory
and when $k$ D-fivebranes collapse to a point we obtain the gauge
group $Sp(k)$ [\Wittentwo]. There is a vector multiplet and a
hypermultiplet whose bosonic components
$X^{AY}$ describe the location of the D-fivebrane which come from the 
$(5,5)$ strings. There are 
also hypermultiplets $h^M_{Am}$ 
coming from the $(5,9)$ strings
[\Wittentwo] although we have set these to zero in the bound state phase. 
In fact it was argued in [\Douglastwo] that the $D=6$, $N=1$ gauge
theory on the D-fivebrane worldsheet has the same
potential as the D-string which is
$$
V = \phi^2(h^2+X^2) \ .
\eqn\sixpotential
$$
Thus in the vacuum $h^{M}_{Am}=X^{AY}=0$, where the D-string is bound to the
D-fivebrane, both the $(5,5)$ and $(5,9)$ strings are massive and so
cannot be turned on. 

Now the D-string appears as  a 1-brane soliton  
corresponding to an instanton of instanton number one
in the space transverse to the string.
The moduli space of $Sp(k)$ instantons with instanton number one 
is $4k+4$ dimensional, which is precisely the number of bosonic degrees of
freedom in the model \confact\ [\Douglastwo]. We therefore interpret the  
target space of the sigma model \confact\ to be the moduli space of
instanton number one self-dual gauge fields on ${\bf R}^4$ with gauge 
group $Sp(k)$. In particular the $b^{AA'}$ fields
describe the centre of the instanton and the $\phi^{A'm}$ 
fields describe its internal structure. The self-dual gauge
field $F^{pq}_{A'mB'n}$ can hence be thought of 
as an ``instanton on instanton moduli space''. 

Let us examine this interpretation in the simplest case, $k=1$, 
which is well known as the
zero-sized gauge fivebrane [\Rey,\CHS]. Here 
$F^{pq}_{mn}$ vanishes everywhere (except at $\phi^{A'm}=0$ where it is 
singular). As mentioned above the $b^{AA'}$ fields simply describe the
position of the centre of the instanton in the D-fivebrane. The four 
$\phi^{A'm}$ then describe the  $Sp(1)\cong S^3$ orientation of the
instanton and its size. 
Thus, even though we started by shrinking 
the spacetime $SO(32)$ instanton to zero size, we recover the full moduli 
space of instantons on the D-fivebrane's worldsheet. 

Note  that in the D-fivebrane metric \metricdef\  the zero size
instantons appear at an infinite distance in the moduli space. This region,
which can be described by a $SU(2)$ WZW model [\CHS,\Reytwo], is a stringy 
version of
the collar neighbourhood in Donaldson's moduli space. Furthermore, since this
WZW model is free of higher order sigma model corrections it should provide
a good description until the massive D-string fields become too light to be
ignored. 

Let us factor out the action of translation in 
the full moduli space of instantons, with instanton number one, 
on ${\bf R}^4$ (i.e. include rotations by the gauge group):
$$
{\cal M}({\bf R}^4) = {\bf R}^4\times{\cal M}_{1}({\bf R}^4) \ ,
\eqn\modspace
$$
where ${\cal M}_{1}({\bf R}^4)$ is the internal structure of the instanton and
${\bf R}^4$
is the location of its centre. The manifold ${\cal M}_{1}({\bf R}^4)$ then 
possesses
a singular point at the origin where the instantons are
shrinking to zero size, which we have already discussed. One may also think
of  infinity as a singular point, where the instanton
is spreading out over all of ${\bf R}^4$. At this point the action of 
translating the instanton centre has a fixed point and one cannot simply
factor out the ${\bf R}^4$ from \modspace. However in the 
metric \metricdef\ both these points are at an 
infinite distance and the space ${\cal M}_{1}({\bf R}^4)$ is smooth. 
Thus the heterotic
sigma model smoothes out the moduli space by moving the singular points to
spatial infinity. 

Another point of note is that the metric which 
appears in the sigma model \confact\ is hyper-K{\"a}hler with torsion, 
whereas with BPS monopole moduli spaces the natural metric (that
is inherited from the Energy density functional)  is simply 
hyper-K{\"a}hler. 
However since we are 
considering string theory on these
manifolds it is  natural to interpret the target space as a 
stringy generalisation of the moduli space metric. (Recently  hyper-K{\"a}hler 
with  torsion geometries have also appeared on  black hole moduli spaces
[\GPS].)  
In a sense the gauge fivebrane metric may be viewed as a 
hyper-K{\"a}hler with torsion blow up of the moduli space. In particular
the $SU(2)$ WZW model describes the sigma model in the region of the
blow up at the origin (see also [\VW] where WZW models make a similar 
appearance). Most often only  
hyper-K{\"a}hler resolutions
of  singular points are considered, which may always be embedded in both the 
heterotic and type II string
theories.  
However the fact that both types of resolution appear in string theory
can be easily seen by T-duality. For example consider a closed string theory on
${\bf R}^4/{\bf Z}_2$. As is well known the twisted sector states act to
blow up the singularity into the hyper-K{\"a}hler Eguchi-Hanson space.
However this space contains a compact killing direction and so can be 
T-dualised into a form of the symmetric fivebrane. This latter structure, 
which is equivalent and moreover  indistinguishable within string
theory to the Eguchi-Hanson space, is hyper-K{\"a}hler with torsion.

For $k\ge2$ the model \confact\ becomes  more complicated. In contrast
to the $k=1$ case the gauge field is non-trivial and the metric is no longer
conformally flat and rotationally invariant. 
In higher dimensions there are $k$ natural
four dimensional coordinate subplanes where the metric
simply reduces to the $k=1$ case and  the four fermion term
vanishes (if the fermions are taken to lie in the tangent space
to the  hyperplane). 
Clearly this corresponds to taking the instanton to lie in some
$SU(2)$ subgroup of $Sp(k)$. In these models the structure is again
hyper-K{\"a}hler with torsion but now the torsion is no longer a closed 
form (because the field strength and hence the chiral anomaly is nonzero) and
these solutions cannot be embedded in type II theories. Although in the next 
section we will see that they may in some sense be thought of as  dual to 
type IIA  hyper-K{\"a}hler structures.

We could generalise the previous construction by considering $l$ D-strings
bound together. In this case we would need to examine the infra-red
fixed point of an $SO(l)\times Sp(k)$ gauge theory in two dimensions 
[\Douglastwo]. The vacua of this theory correspond to the moduli space
of $Sp(k)$ instantons with instanton number $l$ and the low energy effective
action that we would obtain for the D-strings would provide a hyper-K{\" a}hler
with torsion metric on this moduli space. It is natural to suppose that
these fixed point actions would involve  $Sp(k)$ ADHM instantons of instanton 
number $l$ in $4(k+1)l$ dimensions.


\chapter{Type I/Type II Duality}

In the previous section we examined the  sigma model description of the 
bound state of a type
I D-string with $k$ D-fivebranes, let us now consider
an application. To this end suppose we compactify to five dimensions by 
wrapping the D-string/D-fivebrane bound state around ${\bf T}^5$. By 
string/string duality type I string theory on ${\bf T}^5$ is equivalent 
to either 
the type IIA or type IIB  string theories on $K3\times S^1$. 
From the type IIB point of view this bound state
arises as $k$ D-strings wrapped around $S^1$, bound to a single 
D-fivebrane wrapped
around $K3\times S^1$.\foot{Note that string/string duality relates 
electrically charged
states to magnetically charged ones and hence interchanges the roles of
fivebranes and strings.} Whereas in
the IIA string theory the bound state arises from  $k$ D-zerobranes bound to a
single  D-fourbrane wrapped around $K3$. 
In the type II string theory Vafa [\Vafa] has argued that
the low energy effective action for the massless bound state degrees of 
freedom (i.e. the  $(0|4)$ or $(1|5)$ strings [\CM]) is
a hyper-K{\"a}hler $(4,4)$ supersymmetric sigma model with the 
target space
$$
{\cal M}_k(K3) = {(K3)^{k}/ S^{k}} \ .
\eqn\Kmoduli
$$
Here ${\cal M}_k(K3)$ is (up to topological equivalence) the  moduli space of
$SU(2)$ instantons  on $K3$ with instanton number $k$ [\VW]. 
The states which preserve $1/2$ of the 
supersymmetry of the sigma model on ${\cal M}_k(K3)$ can be identified with 
black hole mircostates which preserve
$1/4$ of the spacetime supersymmetry in the five dimensional theory [\SV].
Furthermore by
counting these states, that is states 
composed of a right moving vacuum tensored with an arbitrary number of left 
movers, one can reproduce the Beckenstein-Hawking entropy formula [\SV].
Under string/string duality, the BPS states of this type II sigma model 
must be in a one-to-one 
correspondence with states of the $(1|5)$ strings in the ADHM sigma model  
compactified on ${\bf T}^5$ which preserve all of the  worldsheet  
$(0,4)$ supersymmetry. 

Note that 
the type IIB description arises from the effective action of the D-fivebranes. 
The ground states, which preserve one half of the spacetime supersymmetry, 
correspond to D-string states with no 
D-fivebranes present
and thus do not have an equivalent in the ADHM sigma model constructed above,
since there would then be no massive terms in the action \action.
However, under duality  these states are  mapped to states of the free 
heterotic string 
preserving all of the $(0,8)$ supersymmetry with charge $k$. These are in turn
given by a tensor product of the right moving ground state with an arbitrary
left moving state at level $k$. It was observed in [\VW] that the 
degeneracy of these states, which is the degeneracy of the left moving
modes of a bosonic string at level $k$, is precisely the Euler number of the 
space ${\cal M}_k(K3)$. But this is also the number of vacua of the effective 
sigma model on  ${\cal M}_k(K3)$
in the type II theory and thus  provides a highly non-trivial test of 
string/string duality [\Vafa].

Let us examine the correspondence between the $(1|5)$ strings of the
ADHM model and the $(1|5)$ strings in the type IIB model  more closely.
After compactifying the coordinates $x^1,...,x^5$ on ${\bf T}^5$ the 
ADHM sigma model fields $\phi^{A'm}$ also become periodic and  
the field strength  and metric found
above for  no longer hold. 
However, given our interpretation of the target space
we can still hope to make some comments. First we note that 
since $Sp(k)$ is simply connected there are no instantons
of instanton number one on  ${\bf T}^4$ (see for example [\BvB]). Thus the 
D-string cannot
appear as a single (non-singular) instanton in the transverse space. 
Furthermore it cannot 
appear as an instanton with a higher instanton number since the dimension
of the target space would then be larger than $4k$. Therefore we will
assume that  the D-string can only appear 
as a (singular) zero sized instanton. The moduli space for the $(1|5)$ strings
is then just  a choice of the vacuum transverse to the D-string.
This is analogous to the construction of the collar region in Donaldson's 
moduli space by attaching zero sized instantons  to a flat connection, 
only now there are no moduli corresponding to size and gauge rotations
which would serve to blow up the instanton. 
Thus the target space for the $\phi^{A'm}$ can be identified with the space 
of Wilson lines. 

An arbitrary Wilson line for the group $Sp(k)$ on ${\bf T}^4$ can be 
diagonalised to
$$
W_i = {\rm diag}
(e^{ia_i^1},\ldots,e^{ia_i^k},e^{-ia_i^1},\ldots,e^{-ia_i^k}) \ ,
\eqn\wilsonline
$$
where the $a_i^1,\ldots,a_i^k$ parameterise $k$ copies of the torus 
${\bf T}^4$. However we must also divide out by the action of the
Weyl group [\Wittentwo]. This is generated by $k$  ${\bf Z}_2$ actions 
$a^p_i\rightarrow-a^p_i$ and also an $S^k$ action which permutes the
$p = 1,\ldots,k$ indices. Thus we find the moduli space of Wilson lines to be
$$
{\cal W}_k = {\left({\bf T}^4/{\bf Z_2}\right)^k\over S^k}\ .
\eqn\wlms
$$
Therefore we need to consider the $(1|5)$ fields  
$\phi^{A'm}$, $\chi_-^{Am}$ and 
$\lambda_+^{Ym}$ of the ADHM sigma model on 
the target space ${\cal W}_k$. 
We need to count states
which are in the right moving vacuum with the left movers arbitrary.
Although there is no left moving supersymmetry 
there are equal numbers of fermions and bosons and therefore the left movers 
are in a one-to-one correspondence with the left movers of the $(4,4)$ sigma
model on ${\cal M}_k(K3)$. Next we need to count the number of 
right moving vacua. The right moving
modes have the target space which is an orbifold limit of $(K3)^k/S^k$.
Following the discussion in the previous section  we expect the the sigma 
model metric is a  hyper-K{\"a}hler with
torsion resolution of ${\cal W}_k$. However
the number of vacuum states is determined by the  cohomology of this 
space and it is reasonable to assume that this is the same as a 
hyper-K{\"a}hler resolution of ${\cal W}_k$, i.e. ${\cal M}_k(K3)$.
Thus we find that indeed there are the same number of BPS states as a $(4,4)$
sigma model on ${\cal M}_k(K3)$, in a agreement with the 
type II picture [\Vafa].

Finally we note that the ADHM sigma model \confact\ also contains a free  
action for the massless
$(1|1)$ and $(1|9)$ fields $b^{AA'}$, $\psi_-^{AA'}$ and $\lambda_+^M$. These 
are the same worldsheet fields as a heterotic
$SO(32)$ string on a torus ${\bf T}^4$. In fact these are remnants of the 
full ten dimensional string theory and describe the internal string states 
in the compact dimensions. Under string/string duality 
these are mapped to type IIB $(1|1)$ strings which remain massless 
when the D-strings are bound to  the D-fivebrane where they  form a 
$(4,4)$ sigma model on $K3$.  It is a central observation of heterotic/type II 
duality that the
BPS states of both of these theories  are in a one-to-one correspondence with 
the even self-dual 
charge lattice $\Gamma^{21,5}$. On the heterotic side these arise as the
Narian lattice of internal momentum modes while on the type IIA side these
arise as the wrapping/momentum  modes of the type II D-branes around 
$K3\times S^1$.


\chapter{Comments}

In this paper we have discussed some details of the bound state of a D-string
with $k$ D-fivebranes in type I string theory. This provided an interpretation
of the ADHM construction in higher dimensions within
string theory and was related to the instanton moduli space on $K3$. 
Finally we wish to conclude with  some comments.

The $k=1$ and $k=2$ sigma models discussed in section two may also be 
embedded directly 
into heterotic string theory  in ten dimensions as  $p$-brane solitons 
by simply
adding on $(10-4k)$-dimensional Minkowski space. The $k=1$ case
is the well known gauge fivebrane [\Rey,\CHS] preserving $1/2$ of the $D=10$, 
$N=1$
supersymmetry. Here we have provided this sigma model with another 
interpretation as a string in the moduli space of $SU(2)$ instantons on 
${\bf R}^4$.
The $k=2$ case may be interpreted as a string soliton 
preserving $1/4$
of the $D=10$, $N=1$ supersymmetries [\Lambert] and possessing  
a non-vanishing spacetime gauge field. In fact  the notion of 
self-dual gauge 
fields in higher dimensions contains a variety of examples, each of which 
may be embedded as a heterotic string soliton preserving some supersymmetry. 
In this way one can also obtain 1-branes [\HStwo] and 2-branes [\Ivanova,\GN] 
preserving $1/16$ and $1/8$ of the spacetime supersymmetry respectively 
(there may be other possibilities which have not yet been constructed 
[\CDFN]). To date the meaning
of these solitons has remained obscure. One might think of them as 
intersecting branes but the singularity in the metric is at a single point in 
the transverse space and makes this identification difficult. Furthermore
their mass per unit $p$-volume diverges.
Since we have provided the ADHM 1-brane with a rather 
different role describing bound states this raises the possibility 
that these additional `exotic' branes may also have a similar 
interpretation as effective actions for D-string bound states, possibly after
compactification on a manifold with reduced holonomy.

One final point is that one can apparently pass smoothly
from the spacetime interpretation of the D-string's effective action 
(probe phase) into a 
region where it describes the internal state of the D-string inside 
instanton moduli space (bound phase), although in our case these two
phases are infinitely far from each other. From 
the D-fivebrane point of view the
D-string has ``dissolved'' [\Douglastwo] in the bound state phase. 
On the D-string worldsheet this phenomenon is
reflected by the re-interpretation of the target space as the moduli space of
instantons rather than as spacetime. A similar phenomenon appears 
in the D-brane picture of black holes. In this case 
the point where
a D-string becomes bound to a D-fivebrane ($x^6=x^7=x^8=x^9=0$) 
corresponds to the event horizon
in the low energy supergravity effective action. Classically
the D-string may then further fall beyond the event horizon into the 
singularity. However it was
suggested in [\CM] that in this case the zero-modes of the 
D-string are internal states and do not represent the it's motion in 
physical spacetime, where it remains fixed at the event horizon. 
This blurring of space and moduli space is very intriguing and is perhaps 
pointing to a more fundamental understanding of string theory.

I would like thank P. Howe and H. Verlinde for discussions and the
University of Amsterdam for its hospitality.

\refout

\end